\documentclass[aps,pre,superscriptaddress,floats,showpacs,floatfix,twocolumn,showkeys]{revtex4-1}

\usepackage{bm}
\usepackage{graphicx}
\usepackage{subfigure}
\usepackage{amsmath}
\usepackage{amssymb}
\usepackage{color}
\usepackage[a4paper,left=1.5cm, right=1.3cm, top=3.0cm,bottom=2cm]{geometry}

\begin{document}

\title{Heat and momentum transfer for magnetoconvection in a vertical external magnetic field}
\author{Till Z\"urner}
\affiliation{Institut f\"ur Thermo- und Fluiddynamik, Technische Universit\"at Ilmenau, Postfach 100565, D-98684 Ilmenau, Germany}
\author{Wenjun Liu}
\affiliation{Institut f\"ur Thermo- und Fluiddynamik, Technische Universit\"at Ilmenau, Postfach 100565, D-98684 Ilmenau, Germany}
\author{Dmitry Krasnov}
\affiliation{Institut f\"ur Thermo- und Fluiddynamik, Technische Universit\"at Ilmenau, Postfach 100565, D-98684 Ilmenau, Germany}
\author{J\"org Schumacher}
\affiliation{Institut f\"ur Thermo- und Fluiddynamik, Technische Universit\"at Ilmenau, Postfach 100565, D-98684 Ilmenau, Germany}
\date{September 16, 2016}

\begin{abstract}
The scaling theory of Grossmann and Lohse (J. Fluid Mech. {\bf 407}, 27 (2000)) for the turbulent heat and momentum transfer is 
extended to the magnetoconvection case in the presence of a (strong) vertical magnetic field. The comparison with existing laboratory 
experiments and direct numerical simulations in the quasistatic limit allows to restrict the parameter space to very low Prandtl and 
magnetic Prandtl numbers and thus to reduce the number of unknown parameters in the model. Also included is the Chandrasekhar 
limit for which the outer magnetic induction field ${\bm B}$ is large enough such that convective motion is suppressed and heat is transported 
by diffusion. Our theory identifies four distinct regimes of magnetoconvection which are distinguished by the
strength of the outer magnetic field and the level of turbulence in the flow, respectively.
\end{abstract}
\pacs{47.20.Bp, 47.27.te, 47.65.Cb, 44.25.+f}
\keywords{Rayleigh-B\'{e}nard convection; turbulent convection; magnetoconvection}
\maketitle

\section{Introduction}
One of the central questions in turbulent convection is that of the global transport of heat and momentum as a function of the 
thermal driving and the properties of the working fluid \cite{Kadanoff2001,Ahlers2009,Chilla2012}. In the simplest setting of turbulent 
convection -- the Rayleigh-B\'{e}nard case -- one considers an infinitely extended horizontal layer of fluid which is uniformly heated from below 
and cooled from above. The thermal driving of the turbulent convection in the layer is then established by the temperature difference
between the top and the bottom, $\Delta T=T_{\text{bottom}}-T_{\text{top}}$, and directly proportional to the dimensionless Rayleigh number $Ra$. The properties of the 
working fluid are determined by the Prandtl number $Pr$, defined as the ratio of the kinematic viscosity $\nu$ to the thermal 
diffusivity $\kappa$. Turbulent heat and momentum transfer are quantified by the dimensionless Nusselt, $Nu$, and Reynolds, $Re$, 
numbers, respectively.  In a nutshell, one seeks for $Nu$ and $Re$ being functions of $Ra$ and $Pr$. 

One of the oldest scaling theories that aimed at 
predicting $Nu(Ra)$ at fixed $Pr$ dates back to Malkus \cite{Malkus1954,Malkus1954a} and is based on a marginal stability argument 
for the turbulent mean profiles. More recently, scaling theories by Shraiman and Siggia \cite{Shraiman1990,Siggia1994} and Grossmann and 
Lohse (GL) \cite{Grossmann2000,Grossmann2001} have been developed. The central idea of the GL theory is a decomposition of the thermal and
kinetic energy dissipation into contributions from the bulk and the boundary layers in the vicinity of the plates. These contributions have to be 
weighted with the volume fractions that the boundary layers of the temperature and velocity fields occupy.  The theory is adapted to 
doubly diffusive convection \cite{Yang2015} and horizontal convection \cite{Shishkina2016}.

In astrophysical systems, thermal convection is often tightly coupled to magnetic fields (and rotation) which is known as magnetoconvection
\cite{Weiss2014}. Examples are sunspots in the solar chromosphere \cite{Rempel2011} or the X-ray flaring activity of some young neutron 
stars which are termed magnetars \cite{Castro2008}. Less spectacular, but not less important are numerous industrial 
applications reaching from materials processing, such as crystal growth by the Czochralski method \cite{Series1991} or dendritic solidification 
in alloys \cite{Shevchenko2015}, to fusion technology \cite{Ihli2008}. In case of a strong prescribed magnetic induction ${\bm B}_0$
the secondary magnetic induction ${\bm b}$, which is generated by flow-induced eddy currents, remains very small. While a strong vertical external field can then 
damp and even suppress the convective fluid motion \cite{Chandrasekhar1961}, convection rolls can be stabilized when the magnetic field is applied 
in horizontal direction \cite{Tasaka2016}. From a standard magnetohydrodynamic (MHD) perspective, the turbulence 
of coupled velocity and magnetic fields is then constrained. This regime is known as the quasistatic regime of MHD: the Lorentz force enters the 
momentum equation, the induction equation which describes the temporal evolution of the magnetic induction field ${\bm b}$ is however neglected 
\cite{Davidson2008}.           
 
The aim of the present work is to extend the GL theory of turbulent transport to the case of magnetoconvection. First attempts in this direction have 
been reported by Chakraborty \cite{Chakraborty2008}. He showed that an Ohmic dissipation rate, $\varepsilon_B$, has to be incorporated 
beside the thermal and kinetic energy dissipation rates, $\varepsilon_T$ and $\varepsilon$. One is thus left with eight different regimes of 
boundary-layer- and/or bulk-dominated dissipation rates. Together with free parameters for the viscous boundary layer thickness and a critical Reynolds 
number for the crossover from low to high Prandtl numbers \cite{Grossmann2001}, one ends up with at least ten parameters to fit.  Furthermore, 
dimensionless parameters have to be added that relate the electrical conductivity $\sigma$ either to the kinematic viscosity or the thermal diffusivity 
and quantify the strength of the outer magnetic field. In view to this significant extension of the parameter space, one has to seek for regimes of 
magnetoconvection that can be studied with a reduced set of fit parameters. 

We will therefore restrict the turbulent magnetoconvection to a specific parameter range. 
In view to a comparison with laboratory experiments of magnetoconvection, which are typically conducted in liquid metals, one can restrict 
the Prandtl number range to 
\begin{equation}
Pr=\frac{\nu}{\kappa}\lesssim 10^{-2}\,.
\end{equation}
Also the range of the magnetic Prandtl number $Pm$ can be limited to 
\begin{equation}
Pm=\frac{\nu}{\eta}=\frac{Rm}{Re}\lesssim 10^{-5}\,.
\end{equation}
with the diffusivity of the magnetic induction $\eta=1/(\mu\sigma)$ and $\mu$ being the permeability.
In many laboratory flows the magnetic Reynolds number $Rm$ will thus remain small, $Rm\ll 1$. This regime is
termed the {\em quasistatic} case of magnetohydrodynamics. The magnetic field lines cannot be bended significantly by the fluid 
motion since the magnetic diffusion time scale is very short. 
This excludes some astrophysical applications such as interstellar turbulent gases in which $Pm\gg 1$ \cite{Kulsrud1999}.           
  
Similar to standard GL theory, our predictions have to be fitted to one reference data set. Our adjustment of the free coefficients will be 
based on an experiment by Cioni et al. \cite{Cioni2000} which is to the best of our knowledge the only experiment that was operated at a sufficiently 
high Rayleigh number. Further data records by Burr and M\"uller \cite{Burr2001} and Aurnou and Olson \cite{Aurnou2001} have 
been conducted at smaller Rayleigh numbers 
and will be discussed only briefly. In addition, our own direct numerical simulations of magnetoconvection in the quasistatic regime will be 
included to obtain (at least one) data point with known Reynolds and Nusselt numbers at given Rayleigh, Hartmann (the dimensionless measure for 
magnetic field strength which will be defined in section II) and Prandtl numbers.
 
The outline of the work is as follows. In the next section, the set of magnetoconvective equations of motion is discussed, the characteristic 
scales, dimensionless parameters and dissipation rates are defined. Also the numerical method and a short description of the  data sets will be 
presented. This section is followed by a derivation of the nonlinear equations for $Nu$ and $Re$. Finally the free parameters of the scaling theory 
are fitted to data records. The results are summarized and discussed in brief at the end of the work.   

\section{Equations and parameters}
\subsection{Quasistatic equations of magnetoconvection in Boussinesq approximation}
\begin{figure}[htbp]
\centering
  \includegraphics[width=0.55\columnwidth]{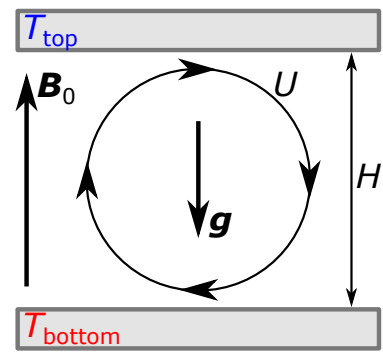}
\caption{(Color online) Magnetoconvection flow. The outer magnetic induction ${\bm B}_0=B_0 {\bm e}_z$, the acceleration due to gravity ${\bm g}=(0,0,-g)$,
the temperature difference and the characteristic large-scale velocity are indicated.}
\label{setup}
\end{figure}

We solve the three-dimensional Boussinesq equations for turbulent magnetoconvection in a rectangular cell of height $H$ and 
side lengths $L$ in the quasistatic limit. The equations for the velocity field ${\bm u}({\bm x},t)$ and the temperature field $T({\bm x},t)$ are 
given by
\begin{align}
\label{ceq}
{\bm \nabla}\cdot {\bm u}&= 0 \,, \\
\label{nseq}\nonumber
\frac{\partial{\bm u}}{\partial t}+({\bm u}\cdot {\bm\nabla}) {\bm u}
  &= -\frac{1}{\rho_0}{\bm \nabla} p+\nu {\bm \nabla}^2 {\bm u} \\
  &\phantom{={}} + g \alpha (T-T_0) {\bm e}_z 
    + \frac{1}{\rho_0}({\bm j}\times {\bm B_0}) \,, \\
\label{pseq}
\frac{\partial T}{\partial t}+({\bm u}\cdot {\bm \nabla})  T
  &= \kappa {\bm \nabla}^2 T \,.
\end{align}
The pressure field is denoted $p({\bm x},t)$, $T_0$ is a reference temperature, $\rho_0$ the constant mass density and ${\bm B_0}=B_0{\bm e}_z$
the magnetic field. The Ohm law for the current density is given by
\begin{align}
\label{cueq}
{\bm j} &= \sigma(-{\bm \nabla}\phi+{\bm u}\times {\bm B}_0)\,,
\end{align}
where the electric potential $\phi$ follows from ${\bm \nabla}\cdot{\bm j}=0$. 
The Rayleigh number is given by 
\begin{equation}
\label{Ra}
Ra=\frac{g\alpha\Delta T H^3}{\nu\kappa}\,, 
\end{equation}
and the Hartmann number by 
\begin{equation}
\label{Ha}
Ha=B_0 H\sqrt{\frac{\sigma}{\rho_0\nu}}=\sqrt{Q}\,.
\end{equation}
The square of $Ha$ is also known as the Chandrasekhar number $Q$.
The variables $g$, $\sigma$ and $\alpha$ denote the acceleration due to gravity, the electrical conductivity and the thermal expansion coefficient, 
respectively.  In a dimensionless form length scales are expressed in units of $H$, velocities in units of the free-fall velocity $U_f=\sqrt{g\alpha\Delta T H}$, 
temperature in units of the outer difference $\Delta T$ and magnetic induction in units of $B_0$. The configuration is sketched in figure \ref{setup}.

\subsection{Direct numerical simulations}
The equations (\ref{ceq}) -- (\ref{cueq}) are solved for a closed Cartesian cell with a second-order finite difference scheme. 
The projection-type scheme is nearly fully conservative. The advection-diffusion equation is solved by semi-implicit scheme 
in which nonlinear terms are treated explicitly and diffusion terms implicitly. The program applies MPI and Open MP.  More 
details are found in \cite{Krasnov2011}. For the fit, we will use two series of direct numerical simulations (DNS):
\begin{itemize}  
\item Series 1: $Ra=10^5$, $Pr=0.025$, $20\le Ha\le 50$. The aspect ratios are $\Gamma_x=L_x/H=1$ and $\Gamma_y=L_y/H=1$.
The grid is non-uniform and contains $128^3$ points. 
\item Series 2: $Ra=10^6$, $Pr=0.025$, $50\le Ha\le 200$. The aspect ratios are $\Gamma_x=L_x/H=1$ and $\Gamma_y=L_y/H=1$.
The grid is non-uniform and contains $128^3$ points. 
\end{itemize}
The boundary conditions are as follows: all walls are electrically insulated walls, i.e. the field lines of the current density are closed inside 
the fluid volume. No-slip boundary conditions hold for the velocity at all walls, the top and bottom walls are additionally isothermal with 
prescribed temperatures $T_{\text{top}}$ and $T_{\text{bottom}} > T_{\text{top}}$, respectively. The side walls are thermally insulated. 
The grid is clustered at the top and bottom walls to resolve the Hartmann layers at the top and bottom and first order quantities. 
We have also performed grid-sensitivity studies to make sure that  the Nusselt number remains constant plane-by-plane (plane at 
constant height $z$).    
 
\subsection{Dissipation rate balances}
In correspondence with classical Rayleigh-B\'{e}nard convection, we can derive exact relations for the mean kinetic energy dissipation rate, 
$\varepsilon$, the mean magnetic dissipation rate, $\varepsilon_B$,
and the mean thermal dissipation rate, $\varepsilon_T$. The fields are defined as
\begin{align}
\label{dissKin}
\varepsilon({\bm x},t) &= 
  \frac{\nu}{2}(\partial_i u_j+\partial_j u_i)^2 \,, \\
\label{dissMagn}
\varepsilon_B({\bm x}, t) &= 
  \frac{\eta}{2\mu\rho_0}(\partial_i b_j-\partial_j b_i)^2 \,, \\
\label{dissTemp}
\varepsilon_T({\bm x},t) &= 
  \kappa (\partial_i T)^2 \,,
\end{align}
with $i, j = x, y, z$.  
Since $\bm B_0$ is constant equation \eqref{dissMagn} contains derivatives of the induced magnetic induction ${\bm b}$ only which arise from 
the eddy currents ${\bm j}$.  In the statistically stationary regime we obtain
\begin{align}
\label{diss1}
\varepsilon+\varepsilon_B &= \frac{\nu^3}{H^4} \frac{(Nu-1)Ra}{Pr^{2}}  \,, \\
\label{diss2}
\varepsilon_T &= \kappa\frac{(\Delta T)^2}{H^2} Nu \,.
\end{align}
The Nusselt number, which quantifies the turbulent heat transfer, is given by 
\begin{equation}
\label{Nusselt}
Nu=1+\frac{H \langle u_z T\rangle}{\kappa\Delta T}\,.
\end{equation}
The global momentum transfer in the magnetoconvective system is quantified by the Reynolds number which is defined as 
\begin{equation}
\label{Reynolds}
Re=\frac{\langle u_i^2\rangle^{1/2} H}{\nu} \,.
\end{equation}
In both definitions $\langle\cdot\rangle$ stands for volume-time average or ensemble average.   
While the thermal balance remains unchanged in comparison to the classical Rayleigh-B\'enard case, the kinetic energy balance differs by the 
addition of $\varepsilon_B$ on the left hand side of Eq. (\ref{diss1}). It results from the Joule dissipation in the presence of a magnetic 
field. For completeness, we also list the definition of the magnetic Reynolds number
\begin{equation}
\label{Reynoldsmag}
Rm=\frac{U H}{\eta}=\mu\sigma U H\,, 
\end{equation}
where $U$ is again given by the root mean square velocity, $U=\langle u_i^2\rangle^{1/2}$.   

\section{Extension of the scaling theory of Grossmann and Lohse} 
\label{GL_ext}
The central idea of the scaling theory is to combine Eqns. (\ref{diss1}) and (\ref{diss2}) with a decomposition 
of dissipation rates into contributions coming from the bulk and the boundary layers (BL) \cite{Grossmann2000,Grossmann2001}. 
The following modifications are made to predict $Nu(Ra, Pr, Ha)$ and $Re(Ra, Pr, Ha)$ for our case at hand: 
\begin{enumerate}
\item[(i)] The relevant boundary layer for the velocity field is the Hartmann layer \cite{Davidson2008} (see also appendix),
\begin{equation}
\label{hartmannBL}
\delta_v=\frac{H}{Ha}\,,
\end{equation}
while the thermal boundary layer thickness remains  $\delta_T=H/(2 Nu)$. Contrary to the original GL theory we do not 
have the free parameter $a$ that appears in the Prandtl-Blasius-type expression $\delta_v=aH/\sqrt{Re}$. 
\item[(ii)] We limit the study to low Prandtl numbers as already mentioned in the introduction. Thus the modification for 
the limit of large Prandtl numbers which has been developed in \cite{Grossmann2001} and the related parameter $Re_c$ are 
not necessary here. This saves a second fit parameter.
\item[(iii)] It is well-known from the linear stability analysis \cite{Chandrasekhar1961} that the critical Rayleigh number $Ra_c$ scales as 
\begin{equation}
Ra_c=\pi^2 Ha^2\,.
\label{Hac}
\end{equation}
If $Ha$ is too big at a given $Ra$, convection is suppressed completely. 
\end{enumerate}
The mean energy dissipation rates will be composed of a boundary layer contribution and a bulk contribution. This results to 
\begin{align}
\label{decompKin}
\varepsilon &= \varepsilon_{bulk} + \varepsilon_{BL} \,, \\
\label{decompMagn}
\varepsilon_B &= \varepsilon_{B,bulk} + \varepsilon_{B,BL} \,, \\
\label{decompTemp}
\varepsilon_T &= \varepsilon_{T,bulk} + \varepsilon_{T,BL} \,. 
\end{align}
The dimensional estimates of the different contributions are given by 
\begin{alignat}{2}
\label{contrScale_kinBulk}
\varepsilon_{bulk} &\sim \frac{U^3}{H} &
  &= \frac{\nu^3}{H^4}Re^3\,,\\
\label{contrScale_kinBL}
\varepsilon_{BL} &\sim \nu\frac{U^2}{\delta_v^2}\,\frac{\delta_v}{H} &
  &= \frac{\nu^3}{H^4}Re^2 Ha\,,\\
\label{contrScale_magnBulk}
\varepsilon_{B,bulk} &\sim \frac{\eta}{\mu\rho_0} \frac{Rm^2 B_0^2}{H^2} &
  &= \frac{\nu^3}{H^4}Re^2Ha^2\,,\\
\label{contrScale_magnBL}
\varepsilon_{B,BL} &\sim \frac{\eta}{\mu\rho_0} \frac{Rm^2 B_0^2}{\delta_v^2} 
     \frac{\delta_v}{H} &
  &= \frac{\nu^3}{H^4}Re^2 Ha^3\,,\\
\label{contrScale_tempBulk}
\varepsilon_{T,bulk} &\sim \frac{(\Delta T)^2U}{H} &
  &= \kappa\frac{(\Delta T)^2}{H^2} Re Pr \,, \\
\label{contrScale_tempBL}
\varepsilon_{T,BL} 
  &\sim \kappa\frac{(\Delta T)^2}{H^2} \sqrt{Re Pr}\,. 
\end{alignat}
The bulk scalings of the kinetic and thermal dissipation rates in \eqref{contrScale_kinBulk} and \eqref{contrScale_tempBulk} are 
the same as in the original GL theory \cite{Grossmann2001}. The argumentation in \cite{Grossmann2001} that leads to \eqref{contrScale_tempBL} 
remains valid for the present case. However, the scaling relation in \eqref{contrScale_kinBL} differs to the original case. 
Instead of the original BL expression $\delta_v^{GL}=aL/\sqrt{Re}$, we insert the Hartmann layer thickness \eqref{hartmannBL}.
For the new estimates in \eqref{contrScale_magnBulk} and \eqref{contrScale_magnBL} we use the definition of 
$\varepsilon_B$ which is given in \eqref{dissMagn} and measure the induced magnetic field ${\bm b}$ in units of $Rm B_0$.

Following Grossmann and Lohse \cite{Grossmann2001}, we introduce interpolation functions to account for changes of 
the scaling laws in different parameter regimes. Once $\delta_T$ becomes smaller than $\delta_v$ the dominant velocity in the 
thermal BL changes from $U$ to $U\delta_T/\delta_v$. This is accounted for by replacing $Re$ with $Re f(x_T)$ in 
\eqref{contrScale_tempBulk} and \eqref{contrScale_tempBL}, where 
\begin{equation}
f(x_T)=\frac{1}{(1+x_T^n)^{1/n}}
\end{equation}
with the argument $x_T=\delta_v/\delta_T=2 Nu/Ha$ and $n=4$. For this interpolation function follows that ${f(x_T\to0)\to 1}$ and 
${f(x_T\to\infty)\to 1/x_T}$. 

Close to the critical Rayleigh number the bulk of the fluid becomes laminar and $\varepsilon_{bulk}$ scales with $Re^2$ 
rather than $Re^3$ as in \eqref{contrScale_kinBulk} for the turbulent regime. This change is modelled by multiplying \eqref{contrScale_kinBulk} 
with 
\begin{align}
g(x^\ast) &= \frac{1}{f(1/x^\ast)}
\end{align}
with the argument $x^\ast = Re/Re^\ast$. From the definition of $f$ follows that $g(x^\ast\to0)\to1/x^\ast$ and 
$g(x^\ast\to\infty)\to1$. The Reynolds number 
$Re^\ast$ marks the range in which the transition from fully developed turbulence to weakly nonlinear time-dependent regime of velocity dynamics
takes place.
Combining all pure scaling laws with the interpolations as just described gives
\begin{align}
\label{diss7}\nonumber
\frac{(Nu-1) Ra}{Pr^{2}Re^2} 
  &= c_1 Re \, g\left(\frac{Re}{Re^\ast}\right) \\
  &\phantom{={}} + c_2 Ha + c_3 Ha^2 + c_4 Ha^3 \,,
\end{align}
\begin{align}
Nu-1 &= c_5 Re Pr f\left(\frac{2Nu}{Ha}\right)
  + c_6 \sqrt{Re Pr f\left(\frac{2Nu}{Ha}\right)}\,.
\label{diss8}
\end{align}
with the seven a priori unknown model parameters $Re^\ast$ and $c_1$ to $c_6$ which have to be determined from a data record. The set of 
implicit equations can then be solved to obtain expressions $Nu (Ra, Ha, Pr)$ and $Re (Ra, Ha, Pr)$. While it is not possible to find a full solution 
analytically, $Re$ can be calculated from \eqref{diss8} as a function of $Nu$, $Ra$, $Ha$ and $Pr$:
\begin{align}
\label{modelRe}
Re &= \frac{\left(\sqrt{c_6^2+4c_5(Nu-1)}-c_6\right)^2}{4c_5^2Pr f\left(\dfrac{2Nu}{Ha}\right)}
\end{align}
Inserting \eqref{modelRe} into \eqref{diss7} gives an equation independent of $Re$. However this new equation cannot be solved analytically for $Nu$.

The stabilizing effect (iii) of large $Ha$ is included here in the following way: assuming we have found an analytical expression $Nu-1=\mathcal N(Ha, Ra, Pr)$ we can enforce the transition to the non-convective regime at $Ra=Ra_c$ by multiplying $\mathcal N$ with 
\begin{align}
h(x_c) &= 1-f(x_c) \,,
\end{align}
where $x_c=Ra/Ra_c$. The function $h(x_c)$ obeys the properties ${h(x_c\to0) \to 0}$ and ${h(x_c\to\infty) \to {1-1/x_c} \to 1}$, which ensures that $Nu\to1$ in the purely diffusive equilibrium. The crossover function transitions smoothly between these two states, so that at $Ra=Ra_c$ we have $h(1)\approx 0.16$ instead of an abrupt jump to zero. Since we cannot determine $\mathcal N$ directly we transform $Nu-1=h(x_c)\mathcal N$ into $(Nu-1)/h(x_c)=\mathcal N$ and in the $Re$-independent equation we replace $Nu-1$ by $(Nu-1)/h(x_c)$. This gives the same result of $Nu=1$ in the non-convective regime once the equation is solved for $Nu$ by numerical methods. Thus the final model equations are \eqref{modelRe} and
\begin{align}
\label{modelNu}\nonumber
\frac{(Nu-1) Ra}{\zeta^2Pr^{2}h(Ra/Ra_c)} 
        &= c_1 \zeta g\left(\frac{\zeta}{Re^\ast}\right) \\
        &\phantom{={}} + c_2  Ha  + c_3 Ha^2 + c_4 Ha^3 \,,
\end{align}
with
\begin{equation}
\zeta = \frac{\left(\sqrt{c_6^2+\dfrac{4c_5(Nu-1)}{h(Ra/Ra_c)}}-c_6\right)^2}
              {4c_5^2 Pr\, f\left(\dfrac{2Nu}{Ha}\right)} \,.
\end{equation}
Now \eqref{modelNu} can be used to determine the seven model parameters $Re^\ast$ and $c_1$ to $c_6$ by fitting the equation to a data set $(Nu, Ra, Ha, Pr)$. However examining \eqref{modelNu} shows, that it is invariant for the following transformations:
\begin{align*}
c_1 &\to \alpha^6 c_1 \,, &
c_2 &\to \alpha^4 c_2 \,, &
c_3 &\to \alpha^4 c_3 \,, \\
c_4 &\to \alpha^4 c_4 \,, &
c_5 &\to \alpha^2 c_5 \,, &
c_6 &\to \alpha c_6 \,, &
Re^\ast &\to Re^\ast/\alpha^2 
\end{align*}
for any $\alpha\in\mathbb R$. This means that the optimal values for the model parameters are ambiguous. To fix this ambiguity we need at least one data point $(Re, Nu, Ra, Ha, Pr)$ which includes the Reynolds number. Then \eqref{modelRe} can be used to calculate $c_6$ as a function of $c_5$:
\begin{align}
\label{c6fix}
c_6 &= \frac{Nu-1}{\sqrt{RePr f\left(\dfrac{2Nu}{Ha}\right)}} 
  - c_5 \sqrt{RePrf\left(\dfrac{2Nu}{Ha}\right)} \,.
\end{align}
With this step the optimal values of all remaining six model parameters $Re^\ast$ and $c_1$ to $c_5$ are unique.
It is absolutely clear that six parameters, which have to be adjusted, is still a large number. Nevertheless, one has to 
keep in mind that the number of free parameters has already been reduced significantly.
We are not aware of any publications that report magnetoconvection data sets including $Re$. Therefore, we are using our own numerical 
simulations to determine data points $(Re, Nu, Ra, Ha, Pr)$ for evaluating \eqref{c6fix}.

\begin{figure*}[t]
\centering
\subfigure[\label{phaseNu}]
  {\includegraphics[width=\textwidth]{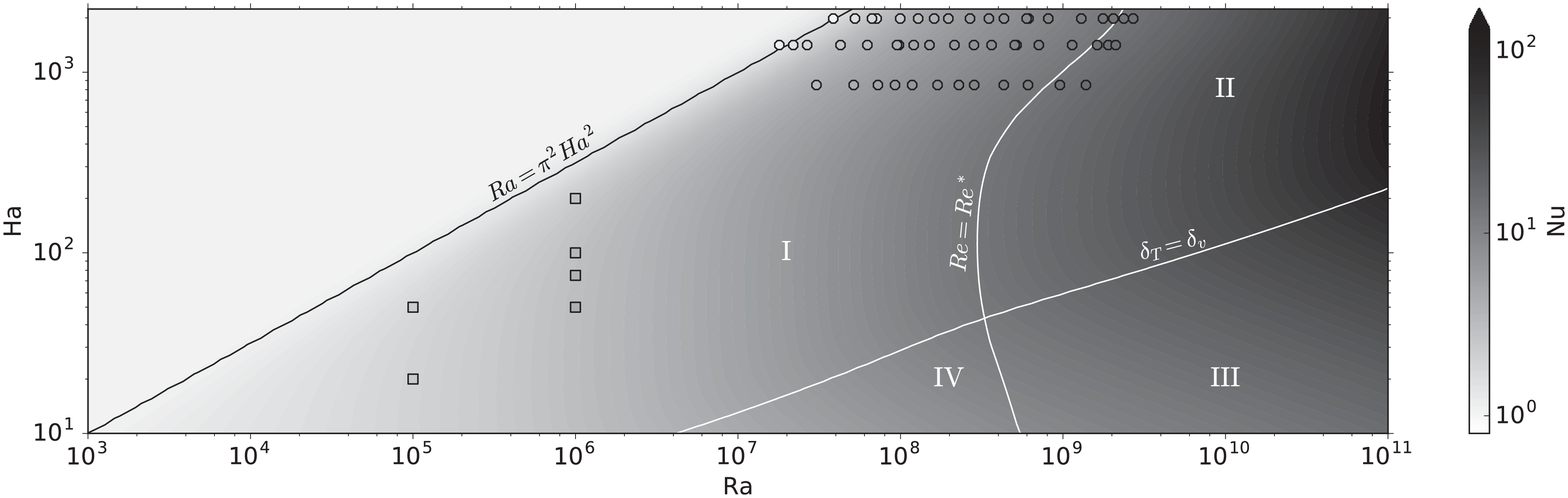}}
\subfigure[\label{phaseRe}] 
  {\includegraphics[width=\textwidth]{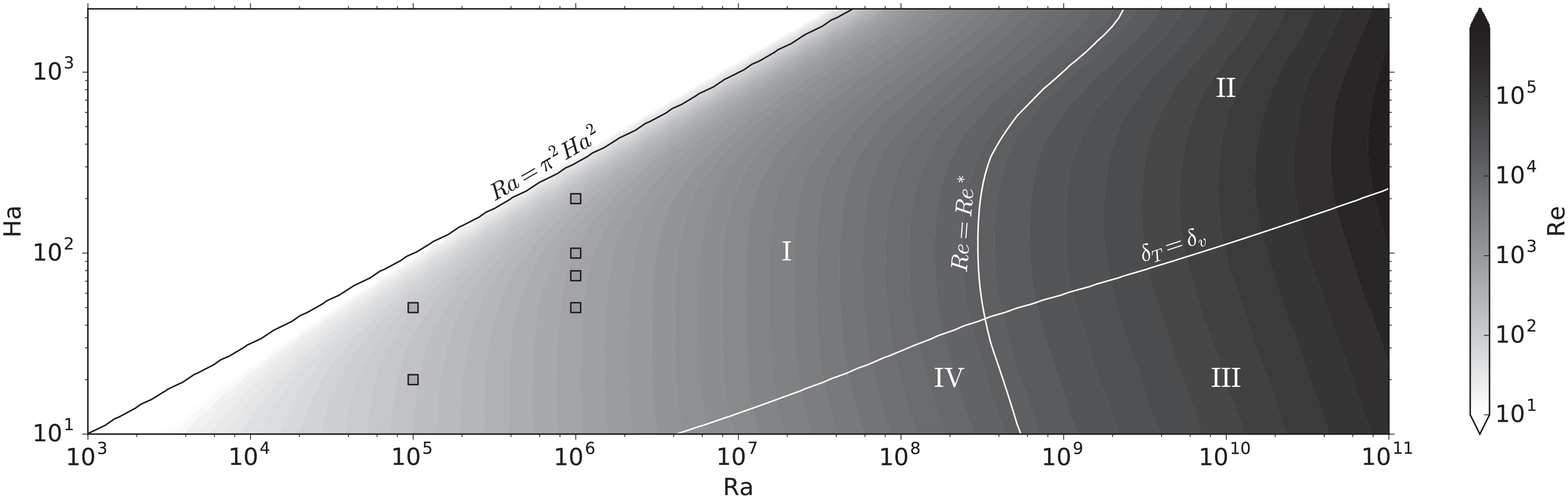}}
\caption{Phase diagrams of \subref{phaseNu} $Nu$ and \subref{phaseRe} $Re$ on the $Ra$-$Ha$-plane for $Pr=0.025$. The symbols represent the data of our numerical simulations (squares) and the experiments by Cioni et. al.~\cite{Cioni2000} (circles) used for fitting the model parameters. The lines in the diagrams mark the position of the crossovers introduced in the model: Below $\delta_T=\delta_v$ the scaling of the thermal BL dissipation changes, $Re=Re^\ast$ marks the transition range from a weakly nonlinear to a fully turbulent bulk flow and $Ra=\pi^2Ha^2$ indicates the onset of convection. Regimes I to IV are marked as described in the text.}
\label{phase}
\end{figure*}
\section{Results}
\label{sectionResult}
Our numerical simulations are used to evaluate \eqref{c6fix}. After substituting  \eqref{c6fix} into \eqref{modelNu}, the resulting equation is fitted to the experimental data of Cioni et al. \cite{Cioni2000} in terms of $Re^\ast$ and $c_1$ to $c_5$, utilizing the Levenberg-Marquardt method \cite{More1978}. The experimental data have been obtained for convection in liquid mercury at a Prandtl number of $Pr=0.025$. Our DNS are conducted at the same Prandtl number. With the known optimal model parameters we can calculate $Nu$ by solving \eqref{modelNu} numerically for given $Ra$, $Ha$ and $Pr$ and subsequently obtain $Re$ from \eqref{modelRe}. The optimal model parameters are
$Re^\ast = 56\,000$, $c_1 = 0.053$,  $c_2 = -2.4$,  $c_3 = 0.014$,   $c_4 = -3.7\times10^{-6}$, and  $c_5 = 0.0038$
From \eqref{c6fix} we get $c_6 = 0.47 $. The $Ra$-$Ha$-phase diagrams for $Nu$ and $Re$ calculated with these 
parameter values for $Pr=0.025$ are shown in figure \ref{phase}. The top panel of the figure shows the magnitude of the 
Nusselt number as a function of $Ha$ and $Ra$. The bottom figure displays the Reynolds number depending on both 
parameters. Also added are the experimental and DNS data. In figure \ref{phase}, we also display the Chandrasekhar limit 
above which $Nu=1$ and $Re=0$.

Furthermore, the line is displayed for which $\delta_v=\delta_T$. Above this line the Hartmann layer thickness will be smaller as the thermal boundary layer thickness. This characteristic line is crossed by a second line that shows $Re=Re^{\ast}$. As mentioned already in section \ref{GL_ext} (see equation (\ref{Hac})), 
on the left side of this line the convection 
flow is not fully developed turbulent, but in weakly nonlinear and time-dependent convection state. All data which are to the right of this line can be considered as fully 
turbulent convection data. It can be seen that only a few data points of \cite{Cioni2000} cross this threshold. The parameter space,  thus, splits into four subregions
by both lines:
\begin{itemize}
\item Region I: weakly nonlinear flow and strong magnetic field 
\item Region II: fully developed turbulent flow and strong magnetic field
\item Region III: fully developed turbulent flow and wea\-ker magnetic field
\item Region IV: weakly nonlinear flow and weaker magnetic field
\end{itemize}
\begin{figure}[t]
\centering
\includegraphics[height=.7\columnwidth]{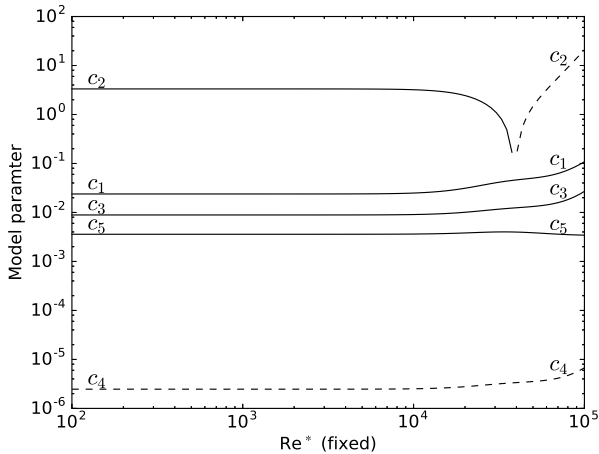}
\caption{Dependence of the coefficients $c_1$ to $c_5$ when fixing the sixth coefficient $Re^{\ast}$.
Negative values of $c_2$ and $c_4$ are indicated by a dashed line.}
\label{fitc2}
\end{figure}

A few words about the quality of the fit should be addressed now. 
First, we mention that the size of the error bars of all fit coefficients (except $c_6$) is of the order of 100~\%. In case of the 
coefficient $c_2$ this error level is even exceeded (see also next paragraphs). This is caused by the sparse record of data points. As can be seen in the 
figure, the data of Cioni et al. \cite{Cioni2000} are collected for three different Hartmann numbers that cover a small range. Also, these data reach only to the 
beginning of regime II. Regimes III and IV do not contain any data points. Stevens 
et al. \cite{Stevens2013} demonstrated in their recent update of GL theory that the uncertainties in the coefficients can be significantly 
reduced when the data cover a wide range of parameters. Furthermore, these three Hartmann numbers are much larger 
than those from our DNS. The additional data by Burr and M\"uller \cite{Burr2001} or by Aurnou and Olson \cite{Aurnou2001} have been conducted close to the 
onset regime of convection. Their experimental data are thus rather in the weakly nonlinear than in the fully turbulent range and will not be used for our study.
\begin{figure*}[t]
\centering
\subfigure[\label{errorNu}]
  {\includegraphics[width=\textwidth]{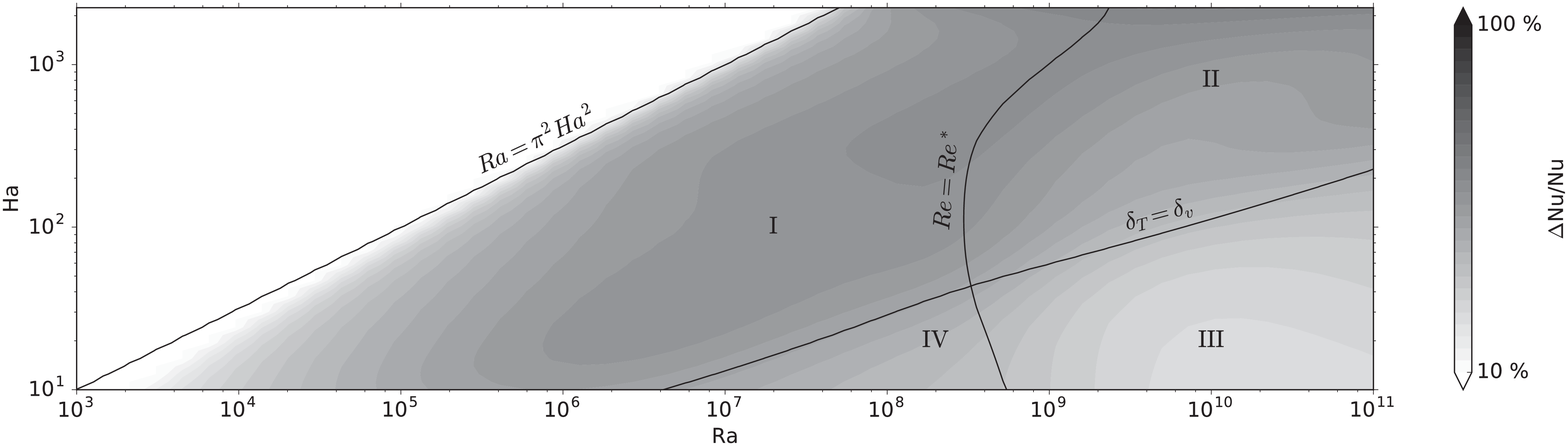}}
\subfigure[\label{errorRe}]
  {\includegraphics[width=\textwidth]{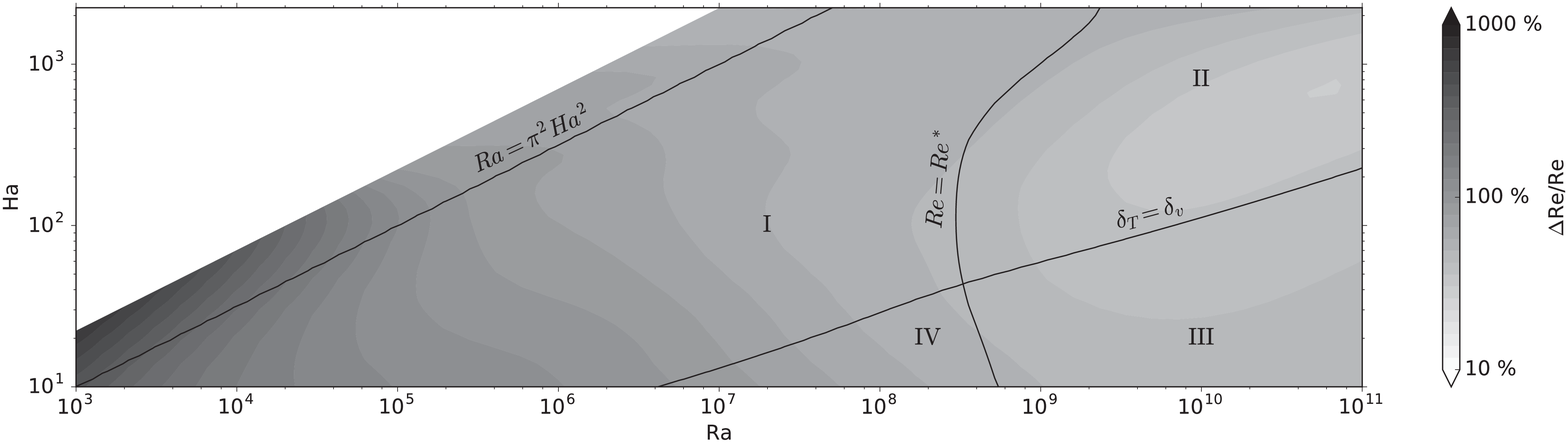}}
\caption{Uncertainty of \subref{errorNu} Nusselt and \subref{errorRe} Reynolds number results in the parameter space. The 
relative errors, $\Delta Nu/Nu$ and $\Delta Re/Re$, are shown in logarithmic scale. The data are obtained by varying the fit coefficients 
$c_1, \dots, c_5$ and $Re^{\ast}$ independently of each other within their error bars. Regimes I-IV and their borders are marked as in 
figure \ref{phase} and the purely conductive regime has been masked.}
\label{phase_err}
\end{figure*}

Secondly, it is observed that two fit coefficients, $c_2$ and $c_4$, are negative. Although $c_4\sim 10^{-6}$  and thus practically zero, the corresponding 
term in (\ref{modelNu}) can give a non-negligible contribution to the scaling due to $Ha^3$. Coefficient $c_2$ with the biggest error bar needs further 
consideration. Figure \ref{fitc2} displays the five coefficients in dependence of a fixed $Re^{\ast}$. To get this figure, we repeated the fits at each fixed value 
of the crossover Reynolds number. It is seen that the results for $c_1$ to $c_5$ are nearly insensitive for $Re^{\ast}\lesssim 2\times 10^4$. Beyond this value, 
coefficient $c_2$ changes sign which is indicated by a dashed line in the plot. The eventual value of $c_2$ falls into a range, where small variations of 
$Re^{\ast}$ cause large changes of $c_2$ (including sign changes).

The magnitude of $Re^{\ast}\sim 5\times 10^4$ in our fit corresponds to a Rayleigh number of $Ra\sim 10^9$. This estimate follows from recent 
numerical  studies in liquid metal convection without magnetic field \cite{Scheel2016}.
It falls thus consistently into the range, for which convection develops into the fully developed turbulent regime which is also known as the hard convective 
turbulence regime \cite{Castaing1989}. At the moment, we can only speculate that the inclusion of more data could lower the value of $Re^{\ast}$ as it is 
expected in low-$Pr$ convection (see e.g. \cite{Mashiko2004,Schumacher2015}).

Thirdly, in order to quantify the impact of the error bars of the fit coefficients on $Nu(Ha, Ra)$ and $Re(Ha, Ra)$, we proceeded as follows. The six coefficients
$c_1, \dots, c_5$ and $Re^{\ast}$ were chosen randomly and statistically independently within their error bars. With these 6-tuples the parameter dependence
$Nu(Ha, Ra)$ and $Re(Ha, Ra)$ is reconstructed for 118 different cases. The superposition of these individual realizations results in an relative error around the 
original value in figure \ref{phase}. The magnitudes of the relative error of both, Nusselt and Reynolds number, are plotted in logarithmic units in figure \ref{phase_err}.  
The relative error of $Nu$ is highest along the border between regime I and II, but does not exceed 40~\%. On the other hand the relative uncertainty of $Re$ rises for smaller $Ra$ and reaches more than 100~\% for $Ra$ below $10^6$.

\section{Summary}
We have presented an extension of the scaling theory of Grossmann and Lohse \cite{Grossmann2000,Grossmann2001}
to a convection layer in the presence of a vertical magnetic field. The discussion is restricted to magnetoconvection at low 
Prandtl and magnetic Prandtl numbers. In this regime the quasistatic approximation is applied that allows a significant reduction of the
number of free parameters in the flow at hand and thus an application of the ideas of GL theory. Below the Chandrasekhar limit four different convection regimes are identified. On the one hand, 
they follow from  the ratio of the Hartmann and thermal boundary layer thicknesses. On the other hand, the regions result from the critical
Reynolds number $Re^{\ast}$, beyond which the convection flow is assumed to be fully turbulent.

In contrast to standard Rayleigh-B\'{e}nard convection, the data base is very small. In fact, there is only one data set from Cioni
and co-workers, that can be used to fit the free parameters. The remaining data \cite{Burr2001,Aurnou2001} fall into a completely different 
section of the parameter plane. In particular, they remain close to the Chandrasekhar limit and cannot be used for turbulent
magnetoconvection. This limits the predictive capabilities of our scaling results and calls for additional experimental data which are planed in 
the near future.

\begin{acknowledgments}
TZ and WL are supported by the Research Training Group GK 1567 on Lorentz Force Velocimetry which is funded 
by the Deutsche Forschungsgemeinschaft. WL is additionally supported by a Fellowship of the China Scholarship Council.
The work of DK is supported by the LIMTECH Research Alliance which is funded by the Helmholtz Association. 
We thank Jonathan Aurnou, Detlef Lohse and in particular Bruno Eckhardt for helpful discussions. 
\end{acknowledgments}

\appendix* \section{Hartmann layer}
The Hartmann problem \cite{Hartmann1937} describes an isothermal pressure-driven plane Poiseuille channel flow subject to a vertical
homogeneous magnetic field (see also \cite{Davidson2008}). Starting point is equation (\ref{nseq}) for $T=T_0$. One seeks a steady solution
$u_x(z)$ in the quasistatic regime. This results in the inhomogeneous differential equation
\begin{equation} 
\rho_0\nu\frac{d^2 u_x(z)}{dz^2} -\sigma B_0^2 u_x(z) = -G\,,
\end{equation} 
with $\partial p/\partial x=-G=\text{const}.$
The Hartmann layer thickness~(\ref{hartmannBL}) arises as the characteristic length scale in the problem and is given by 
\begin{equation} 
\delta_v= \sqrt{\frac{\rho_0\nu}{\sigma B_0^2}}=\frac{H}{Ha}\,.
\end{equation}

\end{document}